\documentclass[twoside]{JINST}
\usepackage{graphicx}
\usepackage{rangecite}
\usepackage{amsmath}
\usepackage[figuresright]{rotating}
\usepackage{afterpage}

\title{Ionization signals from electrons and alpha-particles in mixtures of liquid Argon and Nitrogen - perspectives
on protons for Gamma Resonant Nuclear Absorption applications}

 \author{M.~Zeller\thanks{Corresponding author.}, I.~Badhrees, S.~Delaquis, A.~Ereditato, S.~Janos, I.~Kreslo, M.~Messina, U.~Moser, B.~Rossi  \\
 \llap{}Albert Einstein Center for Fundamental Physics, \\
 Laboratory for High Energy Physics, \\
 University of Bern, \\
 Sidlerstrasse~5,\\
 3012 Bern, Switzerland \\
}

\abstract{In this paper we report on a detailed study of ionization signals produced by Compton electrons and alpha-particles in a Time Projection Chamber (TPC) filled with different mixtures of liquid Argon and Nitrogen. The measurements were carried out with Nitrogen concentrations up to 15$\%$ and a drift electric field in the range 0-50~kV/cm. A prediction for proton ionization signals is made by means of interpolation. This study has been conducted in view of the possible use of liquid $Ar-N_2$ TPCs for the detection of gamma-rays in the resonant band of the Nitrogen absorption spectrum, a promising technology for security and medical applications.}


\begin{document}
\section{Introduction}

The liquid Argon Time Projection Chamber (TPC) is an ionizing particle tracking detector first proposed in 1977 \cite{Rubbia77,Attie}. Due to the higher density of liquids,
the measurement of the linear ionization density dE/dx is more efficient compared to gas TPCs.
This also allows a better identification of the charged particles producing ionization.
The study presented in this paper is part of a comprehensive R\&D program on liquid Argon TPCs that we are conducting in Bern since a few years \cite{Ereditato1,Ereditato2,Ereditato3,nitroTPC,Biagio,Biagio2}.
 
Gamma Nuclear Resonant Absorbtion (GNRA) radiography is an element-specific imaging
technique. In particular, GNRA radiography of Nitrogen is a promising technology for medical
and security applications  \cite{GRA1,GRA2,GRA3,GRA4}. The characteristic feature of 9.17~MeV resonant gamma-conversion in a Nitrogen-rich detector is the production of a proton with an energy of about 1.75 MeV. Non-resonant gammas, however, mostly produce Compton  electrons  with
a nearly flat spectrum from 0 to 9 MeV, assuming that a monoenergetic  9.17 MeV photon beam is used, which corresponds to the resonant absorbtion line in Nitrogen. A significant fraction of $e^+e^-$ pairs is also produced.
Since the width of the resonant line is very narrow (order of 100 eV), it does not seem to be possible to produce an illumination source with such a narrow
emission line. Hence, out-of-resonance gammas are always present. Therefore, the separation of
resonant and non-resonant gammas can be performed if the detector is capable to distinguish 1.75
MeV protons from Compton electrons. In a dense medium (about 1~g/cm$^3$) the tracks left by protons
appear as dot-like clusters with a size of the order of 10 microns, while Compton electrons leave
extended scattered tracks with a few millimeters length and much less ionization density along the
track. Using a tracking detector with spatial resolution of the order of 1 mm, the separation can
efficiently be performed by the combined analysis of ionization density and track topology.

The approach that we are proposing is to use a cryogenic TPC, filled with a
mixture of liquid Argon and Nitrogen. The concentration of Nitrogen is chosen to be high enough
to provide high gamma-conversion efficiency in the resonant band on one hand, and low enough to allow sufficient
charge transport and collection performance on the other. 
In order to prove the feasibility of this approach, the knowledge of the
behaviour of the ionisation charge produced by 1.75~MeV protons in liquid Argon-Nitrogen mixtures with a relatively high (>5\%) content of Nitrogen is needed. In particular, knowing the charge recombination as a function of the electric field in the TPC is very important. 

The first observations of signals in ionization chambers filled with mixture of liquid Argon
and 5~mol~\% Nitrogen date from 1948 \cite{DavidsonLarsh}. Later it was shown that the charge attachment cross section in liquid Argon-Nitrogen mixtures is rather low, so that an efficient transport of the ionization signal in such mixtures is possible \cite{Swan,Swan2,Yoshino,Barabash,Shibamura,Sakai,Sakai2}. In \cite{nitroTPC} we confirmed the reliable detection of ionization signals from electrons with an energy in the range 0.6-1.3 MeV with a TPC filled with a liquid Argon- Nitrogen mixtures (LAr-N TPC), with a Nitrogen content of up to 6~vol~\%.
The present work is focused on the electron-proton separation capability of a TPC filled
with a LAr-N mixture with the Nitrogen content up to 15~vol~\%. Since the injection of 1.75 MeV protons
directly into the TPC medium is an extremely difficult experimental task, we used $^{241}Am$ 
as a source of alpha-particles. Together with Compton electrons from a $^{60}Co$ gamma-source, this
setup provided extreme conditions to study the behaviour of heavy (alpha) and light (electron) particles,
with the protons expected to be in between. The response to protons is hence predicted by
interpolation of alpha and electron data.

\section{Experimental setup, data taking and analysis}
\begin{figure}[htbp]	
\center\includegraphics[angle=0, width=0.5\textwidth]{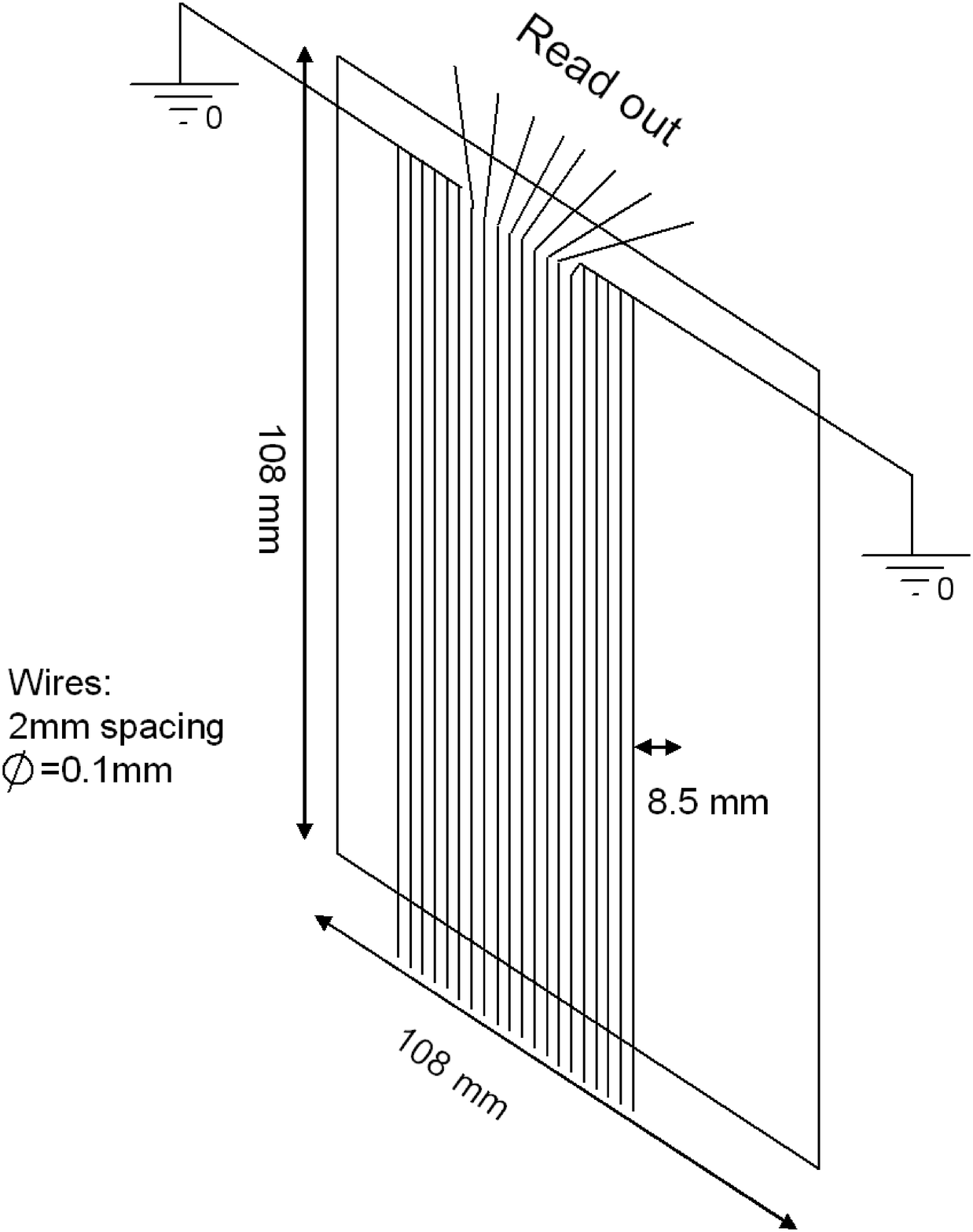}
\caption{Scheme of the TPC. Out of 20 wires only 8 wires located at the center are read out in order to avoid edge effects.}\label{TPC}
\end{figure}

A short drift-gap TPC has been constructed and operated with a relatively high (up to 50 kV/cm) electrostatic drift field.
The chamber has an adjustable drift distance ranging from 3 to 10 mm, 20 tungsten wires with 2 mm spacing, and a length of 103 mm. Out of 20 wires, 8 in the center are read out and the others are connected to ground potential to provide a uniform  electric field in the drift region (see Figure \ref{TPC}). 

The cryogenic system and the TPC charge readout scheme are described in detail in \cite{nitroTPC} where results of measurements of ionization signals from Compton electrons with an energy of up to 1.3~MeV are also presented.

\subsection{Measurement of the Nitrogen concentration in liquid $Ar-N_2$ mixture}
For the studies described in this paper the accuracy on the measurement of the Nitrogen concentration was improved compared to the measurements reported in \cite{nitroTPC}.
Due to the fact that liquid Argon ($\rho$ = 1.394~kg/l) is much denser than liquid Nitrogen ($\rho$ = 0.809~kg/l), the density of the mixture is a good indicator for the Nitrogen concentration. To measure the mixture density we used a tuning fork made of steel, which has a mechanical resonance at a frequency of the order of 2~kHz. The oscillations of the fork are damped by the surrounding medium, which also shifts its resonant frequency.
A tuning fork can be considered as an harmonic oscillator with damping.


The resonant frequency $\omega_{d}$ of such a system is determined by:

\begin{equation}
\label{equ:harmonic2}
 \omega_{d}=\sqrt{\omega^{2}_{0}-\left( \frac{b}{2m} \right) ^2}
\end{equation}

where $\omega_{0}$ is the resonance frequency without damping, $b$ is the damping constant and $m$ is the mass of the oscillating system.

The resonant frequency changes with the damping factor, which increases with the density and the viscosity of the medium. The  temperature affects the resonant frequency of the fork as well. The lower the temperature, the more rigid is the fork, hence increasing the resonance frequency. 
The combination of the above mentioned effects results, to first approximation, in a linear dependence of the resonant frequency on the concentration of Nitrogen in liquid Argon.

\begin{figure}[ht]
\center\includegraphics[width=0.65\textwidth,angle=-90]{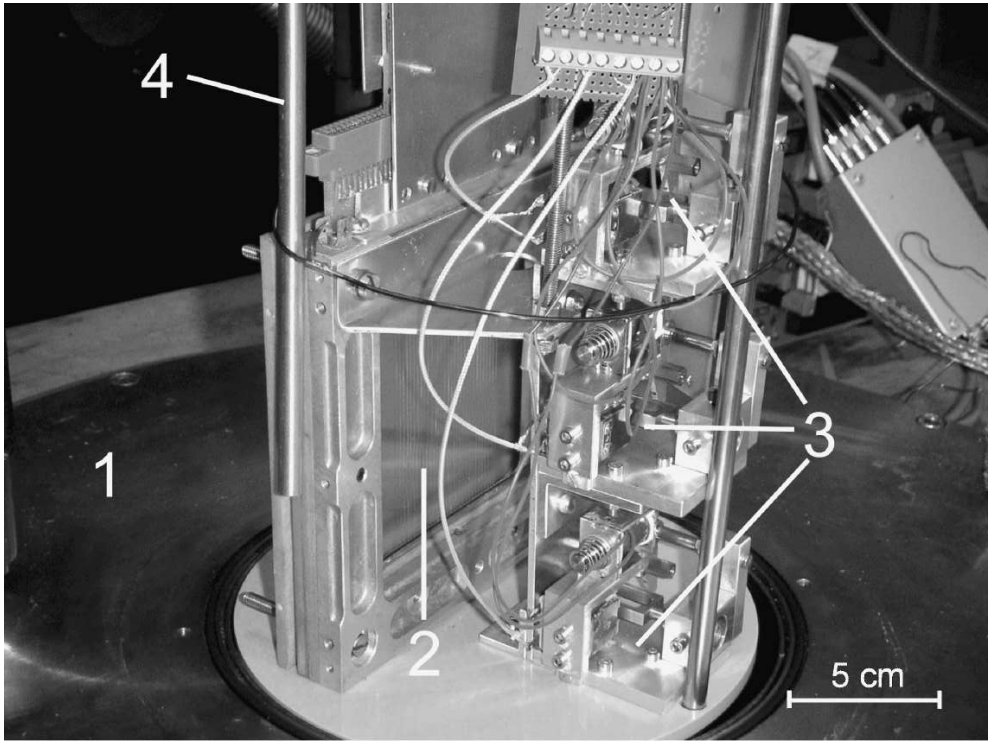}
\caption{Photograph of the detector: 1-dewar top flange, 2-wires of the TPC, 3-resonance based density meters, 4-level meter.}\label{Photowire}
\end{figure}

The experimental setup used for the measurements reported in this paper is shown in Figure \ref{Photowire}, where the newly designed sensors for the Nitrogen concentration in liquid Argon \cite{Marcel} are visible.

\begin{figure}[ht]	
\center\includegraphics[width=0.65\textwidth,angle=-90]{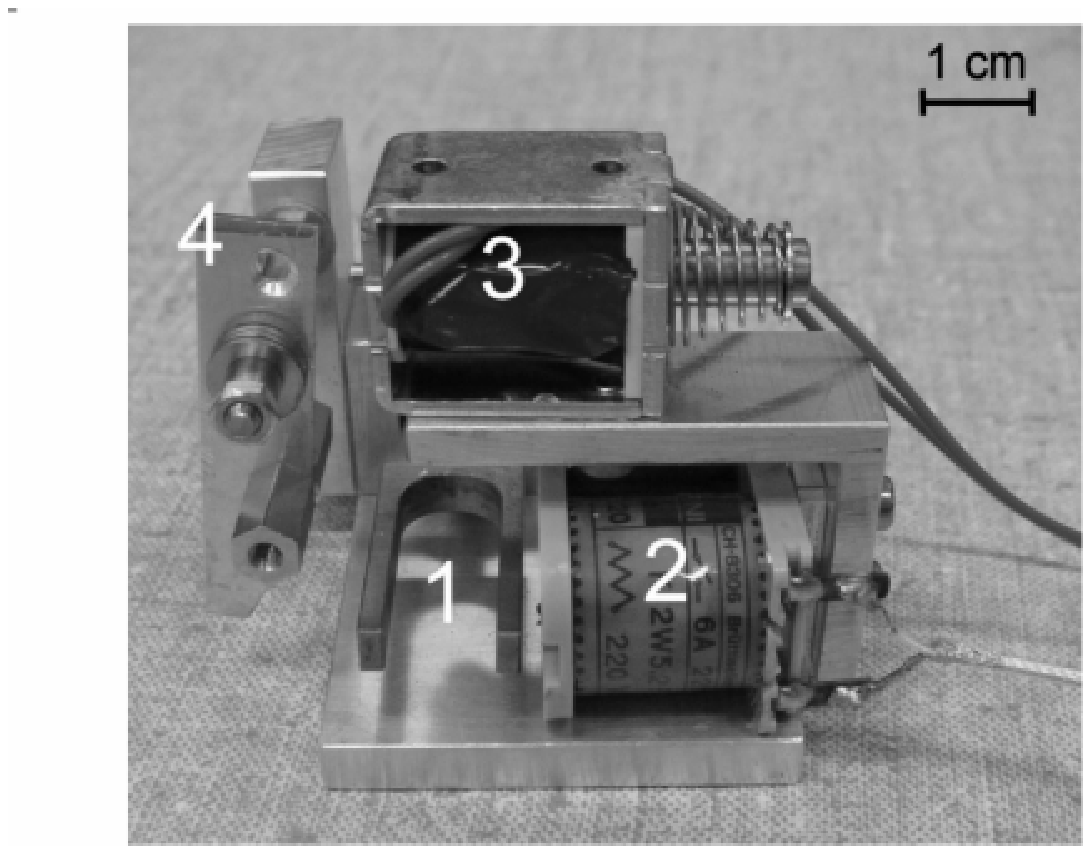}
\caption{The resonance based density meter: 1-fork, 2-receiver coil, 3-lifting magnet, 4-hammer.}\label{fig:Photores1}
\end{figure}

The sensor fork is excited from one side by a miniature hammer driven by an electromagnet. A readout coil with a small permanent magnet in front is placed on the other side of the fork. Figure~\ref{fig:Photores1} shows the  density meter assembly. The oscillation of the fork induces an electric signal in the readout coil. For the monitoring of the possible stratification of the mixture 3 sensors have been built and installed at different depths in the detector vessel.

The signal from the pickup coil is acquired with a dedicated commercial DAQ card\footnote{NI USB-6009 by National Instruments (10 analog inputs, 12 digital I/O, 32-bit counter). }. To extract the oscillation frequency the analysis of the fork oscillations is performed  by a specially conceived LabVIEW program.

\begin{figure}[ht]	
\center\includegraphics[width=0.9\textwidth]{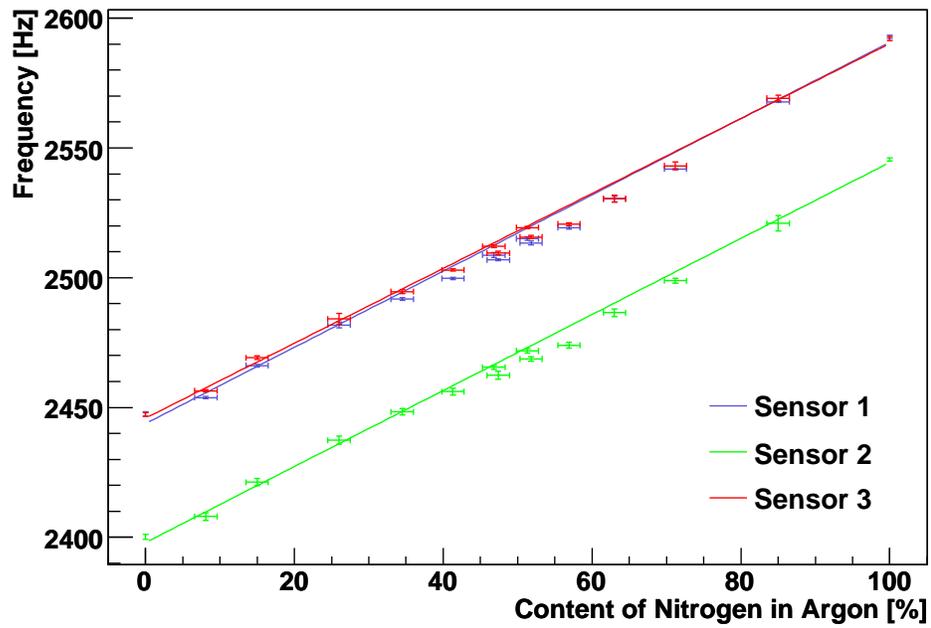}
\caption{Calibration curves of the resonance based density meter (three different sensors), linear fit.}\label{fig:Rescalib}
\end{figure}

The tree density sensors were calibrated with different $Ar-N_2$ mixtures at atmospheric pressure. All three sensors showed a linear dependence in the range from zero to 100$~vol~\%$ of Nitrogen content as presented in Figure \ref{fig:Rescalib}. 
The slight difference in resonance frequency between sensors is due to the fork mechanical tolerances. The statistical accuracy of the Nitrogen concentration measurements was estimated to be $\pm0.1\%$ and the systematic uncertainty of the calibration measurement $\pm0.3\%$ in the concentration range of 0\%-20\%. The sensor is very stable in time, with variations during 24 hours not larger than the statistical error. Throughout this paper the concentration of Nitrogen will be given in 
volume \%, corresponding to the equilibrium temperature at atmospheric pressure (working condition of our TPC).

\begin{figure}[ht]	
\center\includegraphics[width=0.9\textwidth]{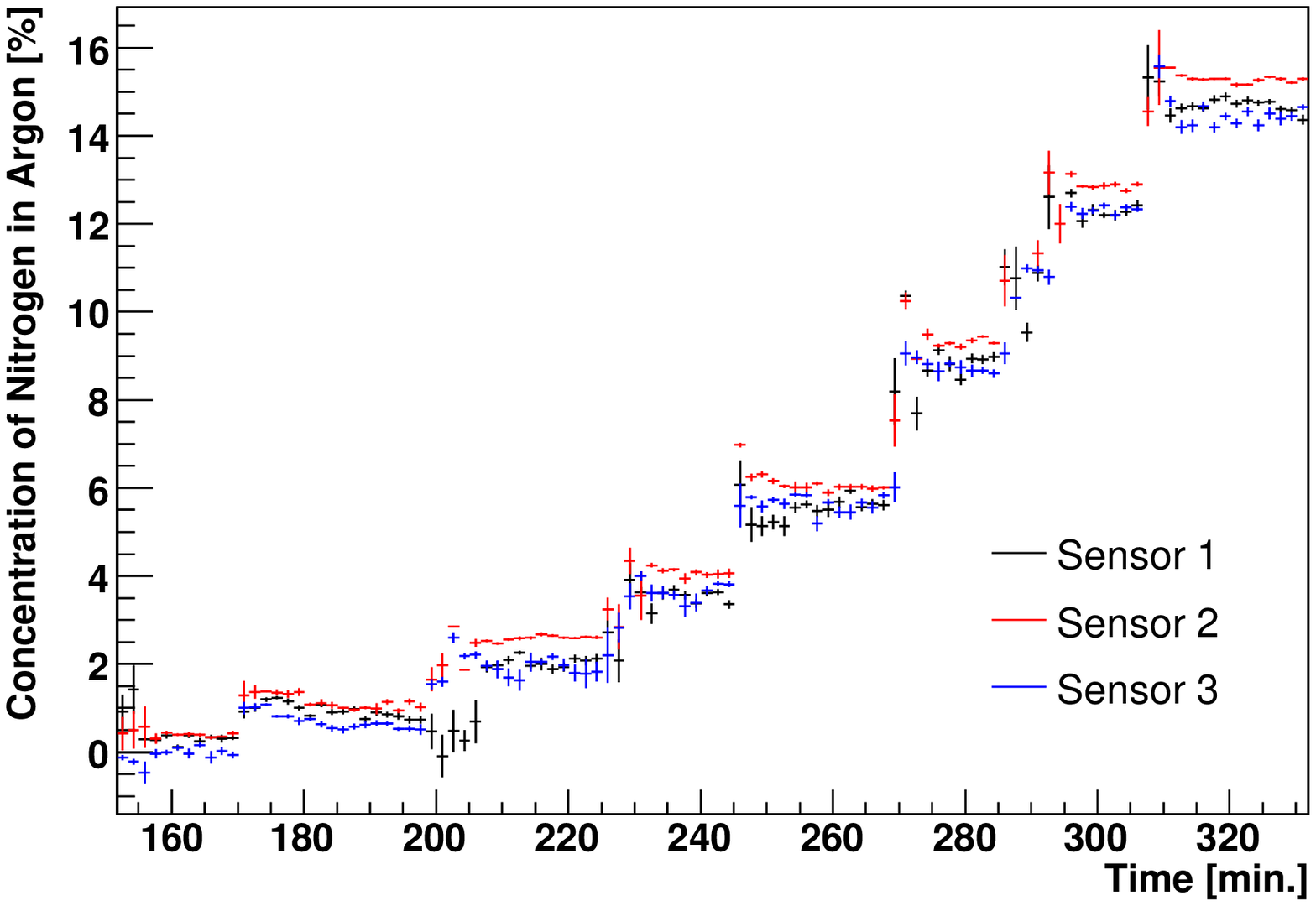}
\caption{Measured concentration of Nitrogen as a function of time during the ionization measurements.}\label{fig:addnitro}
\end{figure}

The detector dewar was initially filled with liquid Argon, purified by an Oxygen-trapping cartridge (for details of the procedure see \cite{nitroTPC}). Then, liquid Nitrogen, purified in the same way, was added.
For the ionization measurements the Nitrogen concentration gradually raised from 0\% to 15\%.  The Nitrogen concentration was permanently monitored by the three sensors located at the top, in the middle, and at the bottom of the detector. The concentration as a function of time is plotted in Figure \ref{fig:addnitro}. For each Nitrogen concentration the ionization signal was acquired for different TPC drift fields. The measurements of the ionization signals started about 10-20~minutes after the Nitrogen addition, in order to let the mixture stabilize. The ionization measurements started when concentration fluctuation went down to 0.1\%. The difference between the concentrations at the top, in the middle and at the the bottom levels of the dewar due to mixture nonuniformity was always within 1.5\%.

Three series of measurements were performed. In order to measure the ionization produced by Compton electrons with energies of about 1 MeV, we used a $^{60}Co$ $\gamma$-radioactive source and a TPC with a drift gap of 8~mm. To characterize the ionization produced by alpha-particles we used an $^{241}Am$ alpha-radioactive source located inside the chamber. In order to estimate drift charge losses due to attachment processes on impurities, a third campaign was conducted with an alpha-source and a smaller TPC drift gap of 5~mm.


The $^{60}Co$ source was placed outside of the dewar. This source emits $\gamma$-rays  of 1.17 and 1.33~MeV. They interact with the detector medium via Compton scattering producing electrons with an energy spectrum ending at the so-called Compton edge. The contribution of the photoeffect is negligible. The energy of the Compton edge for gammas of energy $E_{\gamma}$ is given by $E_c=E_{\gamma}/(1+\frac{m_ec^2}{2E_{\gamma}})$, where $m_e$ is the electron rest mass. These electrons produce electron-ion pairs in liquid Argon, in average one pair per 23.6~eV of deposited energy \cite{Shibamura}. For the 1.33 MeV gammas of~~$^{60}Co$ the maximum Compton electron energy is 1.11~MeV and the total produced charge corresponding to the electrons at this Compton edge is about 7.6~fC. 
We used this value to calibrate our measurements, as described below.

The $^{241}Am$ source was directly placed on the TPC cathode plate. $^{241}Am$ emits alpha-particles with an energy of 5.637~MeV. The source is covered by a protection layer of Au-Pd alloy (5 $\mu$m thick). This reduces the mean energy of the alpha-particles to about $3.3$ MeV (calculated theoretically and confirmed by measurement performed with a scintillator calorimeter), producing in liquid Argon a charge of 22.2 fC (assuming the same 23.6~eV per electron-ion pair as for electrons). The energy  of the alpha-particles after passing the protection layer is smeared according to a Landau distribution. 


\subsection{Calculation of the ionization charge in a TPC}
\label{Calc}

The primary electric charge produced in a drift chamber is reduced by charge recombination during the drift across the anode-cathode gap. The recombination process is well described by the so-called Box model \cite{ThomasImel}, where the ionization charge is assumed to be uniformly distributed in a box volume of linear dimension~$a$ equal to the ionization column diameter, and diffusion effects are neglected. This assumption is appropriate for fields higher than 1~kV/cm, which is the case for the measurements described in this paper. In this model the charge that is left after recombination is given by:

\begin{equation}
q_0=Q_0\frac{1}{\xi} ln(1+\xi),~~~~\xi=\frac{N_0 K_r}{4a^2u_-E}
\label{eq2.2}
\end{equation}

where $Q_0$ is the produced charge, $q_0$ is the charge left after the recombination, $E$ is the applied electric field, $N_0$ is the number of produced electron-ion pairs in the box of linear dimension $a$, $u_-$ is the electron mobility, and $K_r$ is a term related to the recombination cross section. The number of produced electron-ion pairs in the box can be derived from the total ionization charge $Q_0$ and the track length $L$ given by the particle range:

\begin{equation}
N_{0}=\frac{Q_{0}}{e}\frac{a}{L}
\end{equation}

Here, we assume uniform ionization along the track. The remaining charge after recombination can be expressed by:

\begin{equation}
q_{0}=\frac{4au_-eLE}{K_r}ln\left(1+\frac{Q_{0}K_r}{4au_-eLE}\right)
\end{equation}

Introducing $\kappa=\frac{K_r}{4aeLu_-}$ we end up with the following expression:

\begin{equation}
\label{boxmodel}
q_0=\frac{E}{\kappa} ln\left(1+\frac{Q_0\kappa}{E}\right)
\end{equation}

where $Q_0$ corresponds to the total produced charge and $\kappa$ is a parameter depending on the recombination cross section, the dimension of the box, the range of the ionizing particle, and the mobility of the produced charge. This form of the Box model is more appropriate to fit our experimental data than (\ref{eq2.2}). 

After the primary recombination, the spectrum of the collected charge is affected by the following further effects:

\begin{itemize}
\item{Charge loss during the drift.}
 
The source of these losses is the attachment of the drifting electrons to electronegative impurities present in Argon, such as Oxygen, in particular. The collected charge of a TPC after a drift distance $D$ is given by:

\begin{equation}
\label{Dlambda}
q=q_0~e^{-D/\lambda}
\end{equation}

where $q_0$ is the remaining charge after recombination, $D$ is the drift distance, and $\lambda$ is the attenuation length which is a function of the Argon purity and temperature. 

\label{beta}
The probability distribution function of the collected charge $q$ from Compton electrons from monochromatic gamma photons is given by the following composite domain function:

$0<q<q_e\,e^{-D/\lambda}$ 
\begin{equation*}
P(q)=\frac{\lambda}{D} \left[  \frac{8}{3} (e^{D/\lambda}-1) - \frac{8}{3} (e^{2D/\lambda}-1) \frac{q}{E_\gamma} + \frac{14}{9} (e^{3D/\lambda}-1) \frac{q^2}{E_\gamma^2} +\frac{1}{2}  (e^{4D/\lambda}-1)\frac{q^3}{E_\gamma^3} \right];
\end{equation*}

$q_e\,e^{-D/\lambda}<q<q_e$ 
\begin{equation}
P(q)=\frac{\lambda}{D} \left[  \frac{8}{3} \left(\frac{q_e}{q}-1\right)- \frac{8}{3 E_\gamma} \left(\frac{q_e}{q}q_e-q\right)+  \frac{14}{9 E_\gamma^2} \left(\frac{q_e}{q}q_e^2-q^2\right) +\frac{1}{2 E_\gamma^3} \left(\frac{q_e}{q}q_e^3-q^3\right)  \right];
\label{CsA}
\end{equation}

where $D$ is the drift distance between the TPC cathode plate and the wire plane, $\lambda$ is the attenuation length, $E_{\gamma}$ is the energy of the incident $\gamma$, and $q_e$ is the charge left after recombination for electrons with an energy at the end point of the Compton spectrum (Compton edge).
This function is a convolution of the Klein-Nishina formula for Compton scattering with an exponential for charge loss during the drift (\ref{Dlambda}).
In this work the parameter $D/\lambda$ is estimated from independent measurements with alpha particles (see Chapter \ref{Meas}).

\item{The energy resolution of the TPC and the readout system.}

A Monte-Carlo simulation, based on the Klein-Nishina formula for Compton scattering is used to estimate the effect of the detector resolution on the derived value for the Compton edge.  Attachment losses are taken into account assuming $D/\lambda\leq$0.2 (see Chapter \ref{Meas}). The simulated data smeared with the energy resolution are fitted with the function (\ref{CsA}). The discrepancy between the end-point value derived from the fit and the input value from the simulation is then estimated. 

\item{Secondary $\gamma$-rays from Compton scattering.} 

A Monte-Carlo simulation is also used to estimate the effect of secondary scattered gamma-rays from Compton effect on the derived value for the Compton edge. Full volume and the geometry of the detector were simulated, except the dewar walls, since their contribution is estimated to be negligible. 

\end{itemize}

\begin{figure}[!ht]	
\center\includegraphics[width=0.80\textwidth]{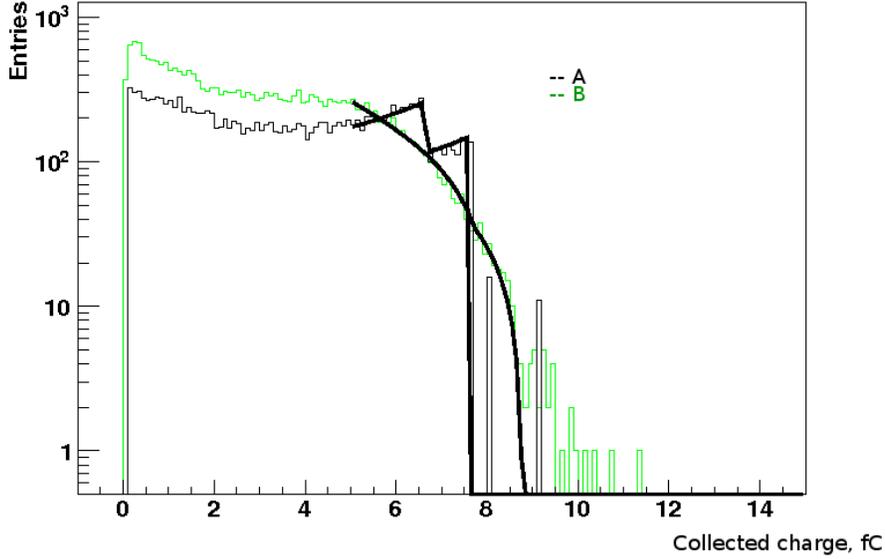}
\caption{Monte-Carlo simulation of the signal amplitude spectrum from Compton electrons with an endpoint corresponding to 7.6~fC of collected charge. The black histogram A represents the energy distribution of Compton electrons from $^{60}Co$ gammas. The green histogram B shows the collected ionization charge after taking into account secondary gammas, attachment losses and detector energy resolution.}\label{fig:sim}
\end{figure}

The endpoint value of the Compton spectrum derived from the fit (\ref{CsA}) is found to be practically insensitive to the effect of secondary $\gamma$-ray interactions, since the effect contributes just to the low energy part of the spectrum. The attachment losses at the purity level of our setup are found to be negligible as well (see Chapter \ref{Meas}). However, the endpoint value is considerably affected by the energy resolution of the detector. The combined effect is illustrated in Figure \ref{fig:sim}. For the case shown in this Figure, the value of the Compton edge derived from the fit is off by about 18\% from the true value and has to be corrected by this factor. The correction factor is calculated as the function of the detector energy resolution, and applied to the fitted data points.

\subsection{Measurement of the ionization produced by Compton electrons and alpha-particles}
\label{Meas}
The main goal of the conducted measurements is to find out parameters of the Box model (\ref{boxmodel}) and their dependence on the Nitrogen concentration for alpha-particles and for Compton electrons. For this task the calibration of the detector (establishing the
correspondence between millivolts of the pulse amplitude on ADC and femto-Coulombs of the charge collected by the wire) is necessary. As explained in the previous Chapter, the detector energy resolution and the drift losses must be measured experimentally in order to be taken into account for detector calibration, while the secondary gamma effect can be corrected for on the basis of Monte-Carlo simulation only.

The ionization charge collected by the TPC wires is measured using fast charge-sensitive amplifiers. The TPC charge readout scheme is described in detail in \cite{nitroTPC}.

\begin{itemize}
\item{Measurement of the energy resolution of the TPC and the readout system.}

In order to measure the energy resolution of our TPC we used an alpha-source based on the $^{241}Am$ isotope. The energy spectrum of the  $^{241}Am$ has a major line ($\approx85\%$)  peaked at 5.486~MeV. A thin (<1 $\mu$m) $^{241}Am$ layer is deposited in a Silver substrate and sealed with a 5 $\mu$m thick protection layer of Au-Pd alloy. The energy spectrum of the source is calculated analytically from the energy loss across the 5 $\mu$m thick protection layer, using Landau distribution. The mean energy is found to be 3.3~MeV. The spectrum is approximated by a Gaussian with $\sigma=0.16$~MeV. For a cross-check the mean energy is measured by the scintillator calorimeter and found to be in a good agreement with calculations.

An alpha-particle of this energy stops in liquid Argon within few tens of microns, so the signal from ionization is always collected by only one TPC wire. This makes the detailed distribution of the energy deposition along the track irrelevant. Experimentally we obtain the ionization charge distributions shown in Figure \ref{fig:Aspec}. These distributions are fitted with a Gaussian function. The mean value of the fit corresponds to the collected charge, the width is determined by the energy resolution of the detector and the energy distribution of the source. The energy resolution improves with increasing drift field and gets worse by adding Nitrogen. 

\begin{figure}[ht]	
\center\includegraphics[width=0.89\textwidth]{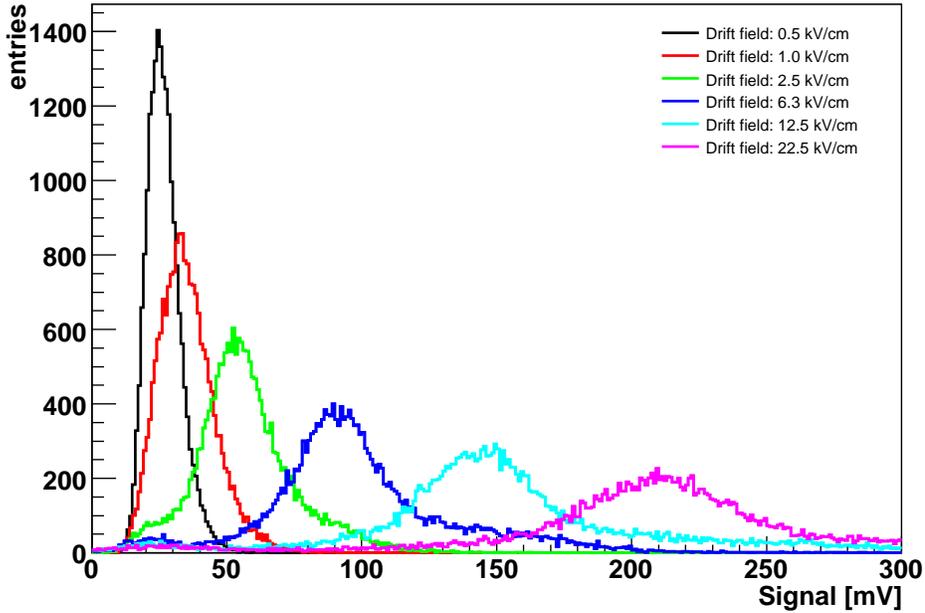}
\caption{Measured signal amplitude spectra from ionization induced by alpha-particles from $^{241}Am$ for different drift fields. The TPC is filled with pure liquid Argon.}\label{fig:Aspec}
\end{figure}

The energy resolution of the detector was derived by subtracting in squares the RMS of the source energy spectrum from the RMS of the ionization signal spectrum from $^{241}Am$ alpha-particles. In Figure \ref{fig:EResG5} the resulting energy resolution of the detector is shown. 

\begin{figure}[ht]	
\center\includegraphics[width=0.89\textwidth]{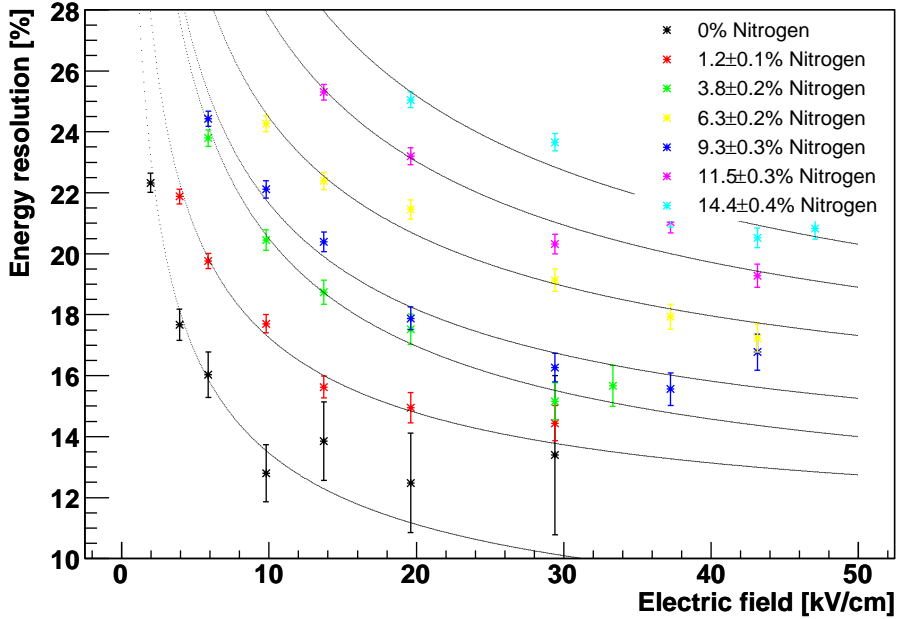}
\caption{TPC energy resolution for different drift-field intensities and different Nitrogen concentrations. The drift distance is 5.1 mm. Data points are fitted with the $A+B/\sqrt{Q}$ function, where Q is the collected ionization charge.}\label{fig:EResG5}
\end{figure}

\item{Measurement of charge loss during the drift.}

Alpha-particles data taken at two different drift lengths have been studied in order to estimate the effect of charge loss due to electronegative impurities. In Figures \ref{fig:AlphaG8} and \ref{fig:AlphaG5} the collected charge is shown as a function of the electric field for different Nitrogen concentrations and drift length 7.9 mm  and 5.1 mm respectively. By comparing the two measurements no significant difference was observed. This means that the attachment losses can be neglected for the drift distances of this order and the obtained purity of the Argon and Nitrogen.  For pure Argon this is also suggested by results published in \cite{Biagio}, where a similar technique of the liquid Argon purification is used. For the mixture of $Ar-N_2$ we set the upper limit for the ratio $D/\lambda \leq 0.2$ estimated from the charge measurement error, comparing measurements with two drift lengths.

\begin{figure}[ht]	
\center\includegraphics[width=0.89\textwidth]{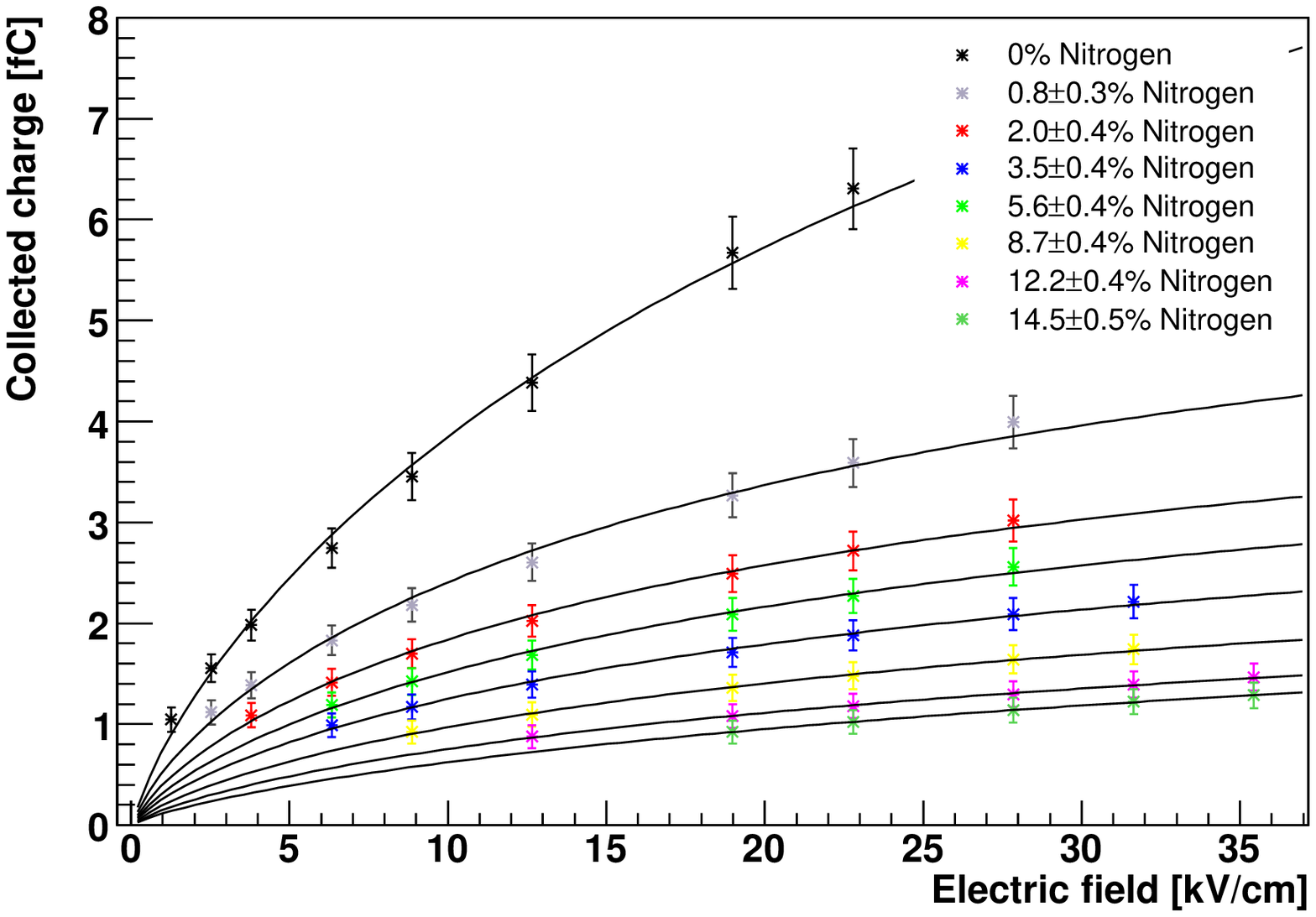}
\caption{Collected charge produced by an $^{241}Am$ source for different drift-field intensities and Nitrogen concentrations. The $^{241}Am$ source is placed directly at the cathode plate of the TPC, so that the drift distance is 7.9 mm. The calibration procedure described in the text is applied. Points are fitted with the Box model.}\label{fig:AlphaG8}
\end{figure}

\begin{figure}[ht]	
\center\includegraphics[width=0.89\textwidth]{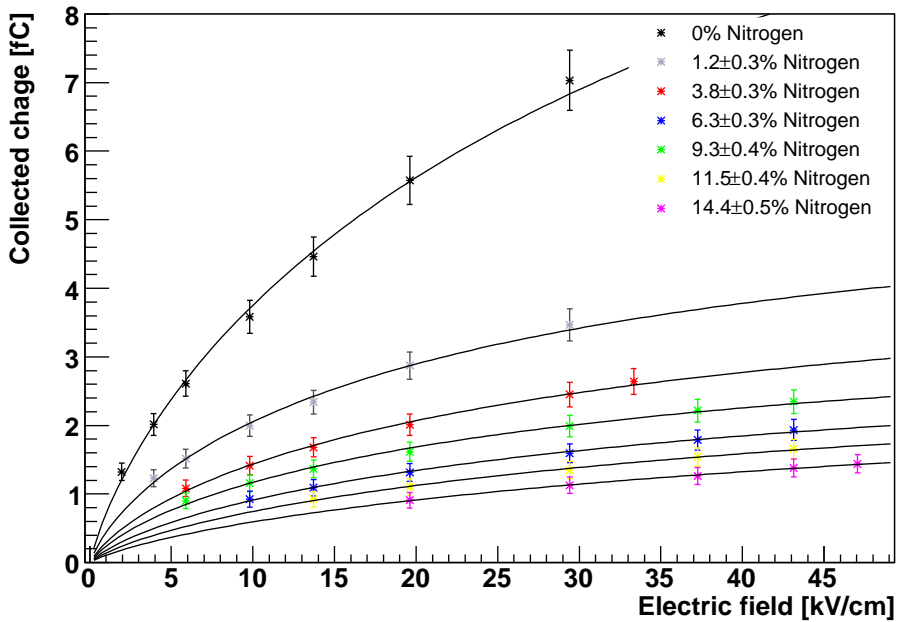}
\caption{Same plot as in Figure 9, but for a drift distance of 5.1 mm.}\label{fig:AlphaG5}
\end{figure}

\item{Detector calibration.}

In order to establish the relationship between the electric pulse read out by the preamplifiers and the charge collected by the TPC wires, the following effect must be taken into account. The ionization density of alpha-particles is higher compared to minimum ionizing particles (MIP). This effect strongly enhances the recombination losses even for fields up to 47~kV/cm, the limit reached in our measurements. On the other hand, the W-value (fraction of the deposited energy needed to create one electron-ion pair) is well known for MIPs, but not so precisely measured for heavily ionizing particles. This suggests to use Compton electrons for calibration, rather than alpha-particles. Comparing the end-point value derived from the fit of the signal distribution in millivolts (corrected for the effects, desribed above) with the calculated theoretical value of the electron energy corresponding to the Compton edge allows to derive the conversion factor from mV of registered pulse to fC of collected charge. This conversion factor, measured for electrons in pure Argon, is then applied to the measured data for both alphas and electrons in $Ar-N_2$ mixtures.
\end{itemize}

To produce Compton electrons we used a $^{60}Co$ gamma source placed outside of the dewar. The distributions of ionization signals from Compton electrons are fitted with the probability distribution function (\ref{CsA}).  Since $^{60}Co$ emits photons of 1.17 MeV and 1.33 MeV with equal intensities, the fitting function is represented by the sum of two functions of the form (\ref{CsA}) with corresponding values of $E_\gamma$. 
The fit allows to derive the collected charge, corresponding to the endpoints of both the 1.17~MeV and the 1.33~MeV photon Compton spectra. The measured energy resolution for each drift field and Nitrogen concentration and the limiting value for $D/\lambda \leq 0.2$ are used to derive the correction factor for a corresponding Compton edge as described in Chapter \ref{Calc}.

\begin{figure}[ht]	
\center\includegraphics[width=0.85\textwidth]{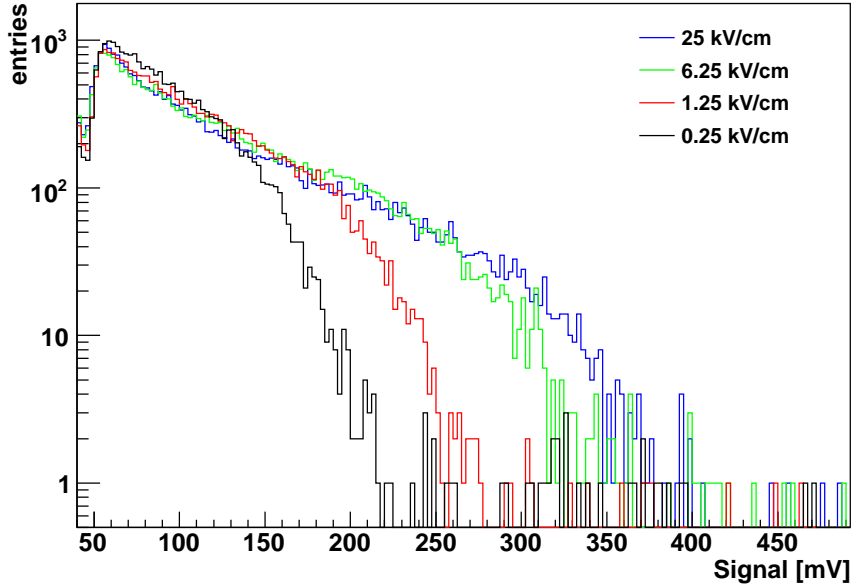}
\caption{Measured spectra of the ionization charge from  $^{60}Co$ Compton electrons for different drift fields. The TPC was filled with pure liquid Argon.}\label{fig:Gspec}
\end{figure}

\begin{figure}[ht]	
\center\includegraphics[width=0.85\textwidth]{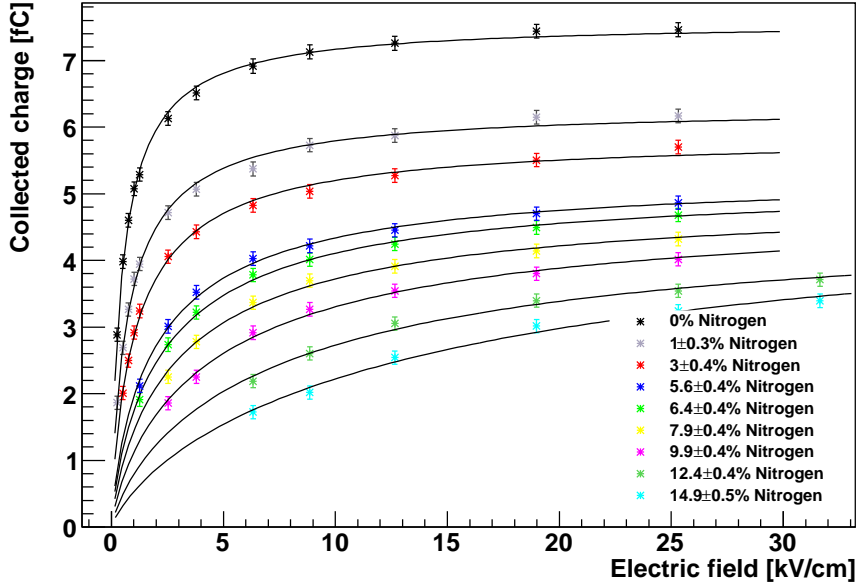}
\caption{Collected charge corresponding to the Compton edge from $^{60}Co$ for different drift field intensities and different Nitrogen concentrations. Experimental data are fitted with the "Box model".}\label{fig:GammaG8}
\end{figure}

In Figure \ref{fig:Gspec}, measured Compton spectra at different electric field intensities are shown.
Figure \ref{fig:GammaG8} shows the collected charge derived from the fitted end points of the Compton spectra at different electric field intensities and different Nitrogen concentrations. The  error of the endpoint value is found to be in the range of 1.5\% to 6\%. 

In order to compare the charge recombination for Compton electrons and alpha-particles we used the Box recombination model in the form (\ref{boxmodel}). The experimental curves in Figures \ref{fig:AlphaG8},\ref{fig:AlphaG5} and \ref{fig:GammaG8} are fitted with this function. This allows to determine the parameters $\kappa$ and $Q_0$ as a function of the Nitrogen concentration in the mixture. The following conversion is then made, using for $L$ values from the CSDA range tables~\cite{CSDA}:

\begin{equation}
\frac {au_-}{K_r}=\frac{1}{4eL\kappa},~~~~W=\frac{e*E_{p}}{Q_0},
\end{equation}

where $W$ is the energy needed to create one electron-ion pair in the $Ar-N_2$ mixture and $E_{p}$ is the energy of the ionizing particle, namely the mean energy for alpha-particles, and the Compton edge energy for electrons. 

\begin{figure}[ht]	
\center\includegraphics[width=0.99\textwidth]{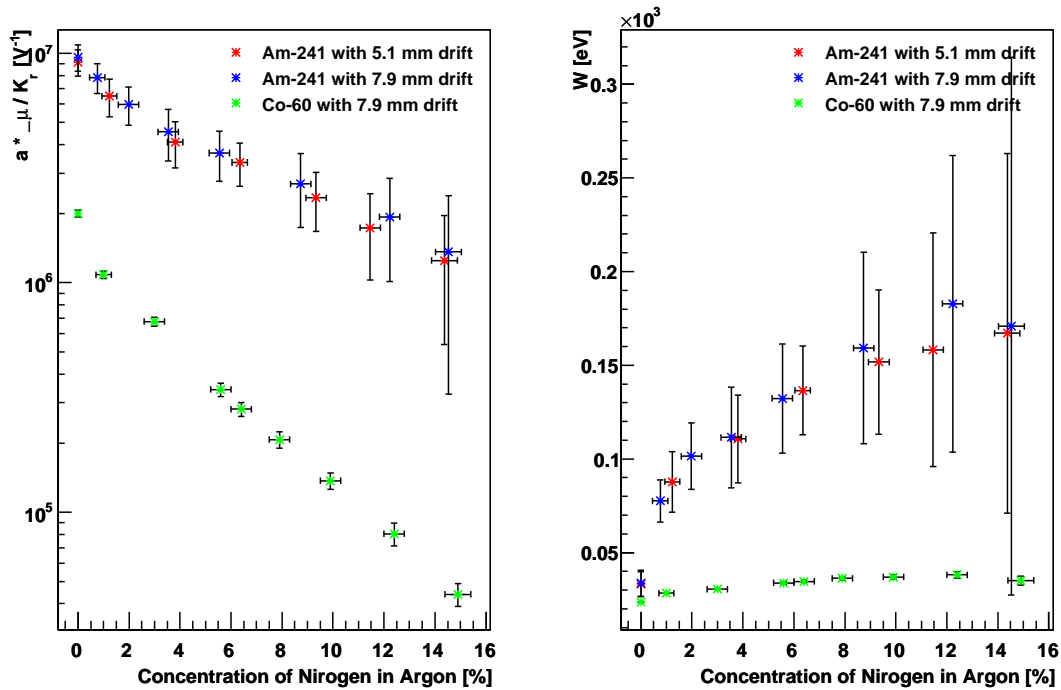}
\caption{Left: $\frac {au_-}{K_r}$ parameter for the alpha particles and electrons as a function of the Nitrogen concentration. Right: measured data of the W value as a function of the Nitrogen concentration.}
\label{fig:Results}
\end {figure}

In Figure \ref{fig:Results} the parameters $\frac{au_-}{K_r}$ and $W$ are shown as a function of the Nitrogen concentration in the mixture. 

The parameter that characterizes the recombination factor $\frac{au_-}{K_r}$ shows a large difference for alpha-particles and electrons at 0\% Nitrogen concentrations (factor of about 10). This reflects the difference in the ionization column diameter $a$, which is larger for heavily ionizing alpha-particles, since their tracks are "dressed" with $\delta$-electrons.

The drop of  $\frac{au_-}{K_r}$ with the increasing Nitrogen content can be partially attributed to the drop of electron mobility $u_-$, as demonstrated in \cite{Barabash}. However, the different slopes of these curves for electrons and alpha-particles suggest also a variation in the track diameter $a$.

At 0\% Nitrogen concentration, $W$ for both alpha-particles and electrons is equal to well known value of 23.6~eV \cite{Shibamura} within the statistical error. Therefore, the energy needed to create an electron-ion pair is similar in both cases. This is an indirect confirmation of the calibration accuracy, since the conversion from a signal in mV to a charge in fC is made only for the electron data and then applied to the alpha-particle data. 

Increasing the Nitrogen concentration leads to quenching of the ionization due to faster electron thermalization.  The thermalization is strongly enhanced by the presence of the rotational and vibrational energy levels of molecular Nitrogen below 10~eV. Hence, the $W$ value increases with the Nitrogen content. However, the effect seems to be much stronger for heavily ionizing alpha-particles than for electrons. A similar difference was observed in \cite{Vuillemin} for a mixture of liquid Argon with another molecular dopant, Ethylene. Quenching seems to be more effective in the region with high charge density along the particle track. The explanation could be in the multi-body interaction kinematics in the region of high ionization density for alpha-particles, when part of the energy transfered to molecular dopant can leak to the free electrons produced by ionization.

\section{Prospects for GRNA applications} 

In the resonant absorption of 9.17~MeV gamma-rays by $^{14}N$ nuclei, the latter decay into a proton and a $^{13}C$ nucleus. The proton has an energy of 1.75~MeV, and leaves in the liquid Argon an ionization track about 10 $\mu$m long. 
In order to experimentally study the ionization yield from such protons in the liquid $Ar-N_2$ mixtures,
one would need a proton source, that can deliver 1.75~MeV protons directly into the TPC drift volume. To our knowledge such a radiation source does not exist at present. The only way such protons can be produced in the TPC is the Gamma Nuclear Resonant Absorbtion. A monochromatic source of gamma photons with a spectral width comparable to the GRNA resonance width (order of 100~eV) is also not easily available. The only feasible way to investigate the ionization behaviour of the protons is the extrapolation from electrons (lighter particles) on one hand and alpha-particles (heavier particles) on the other.

Using the results we have obtained from the measurement of the ionization and recombination parameters for electrons and alpha-particles, we can extrapolate them to the case of GRNA protons, and predict the amount of collected charge for different Nitrogen concentrations and electric field values. This allows a pre-optimization of the operation conditions for future GRNA radiography prototypes. 

\begin{figure}[ht]	
\center\includegraphics[width=0.99\textwidth]{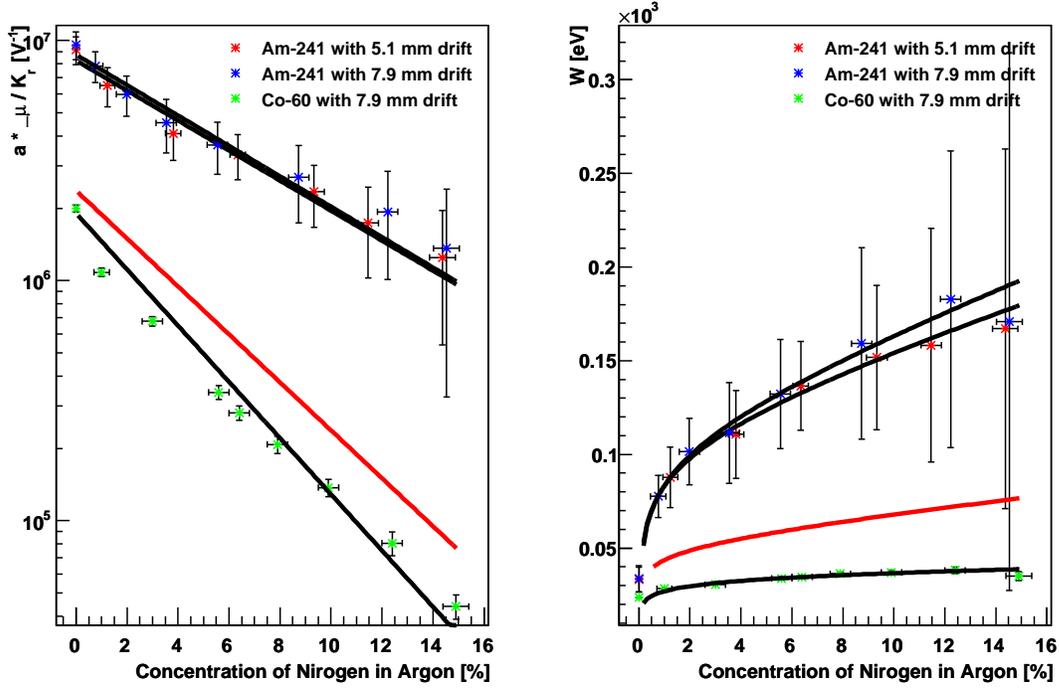}
\caption{Left: $\frac{au_-}{K_r}$ parameter as a function of the Nitrogen concentration, fitted by an exponential function. Right: measured W values as a function of the Nitrogen content, fitted by a linear plus logarithmic function. Red line: interpolated values for protons.}\label{fig:Results_P}
\end {figure}

In order to derive an empirical analytic dependence of the $W$ and $\kappa$ parameters on the Nitrogen concentration, we fitted the curves of Figure \ref{fig:Results} with exponential and logarithmic functions, respectively. Although this fit is not satisfactory below 1\% of Nitrogen content, at higher values it matches well the experimental data. To derive a similar dependences for protons we only considered the first order of the dependences of the $W$ and $\kappa$ parameters on the ionization density dE/dx. This rough approximation assumes no detailed knowledge
of the dependences of the recombination enhancement and the ionization quenching on the ionization density. The systematic study of these dependences is out of the scope of this paper. The results are shown in Figure \ref{fig:Results_P}. The red line represents the expected curves for protons of 1.75~MeV.

\begin{figure}[ht]	
\center\includegraphics[width=0.85\textwidth]{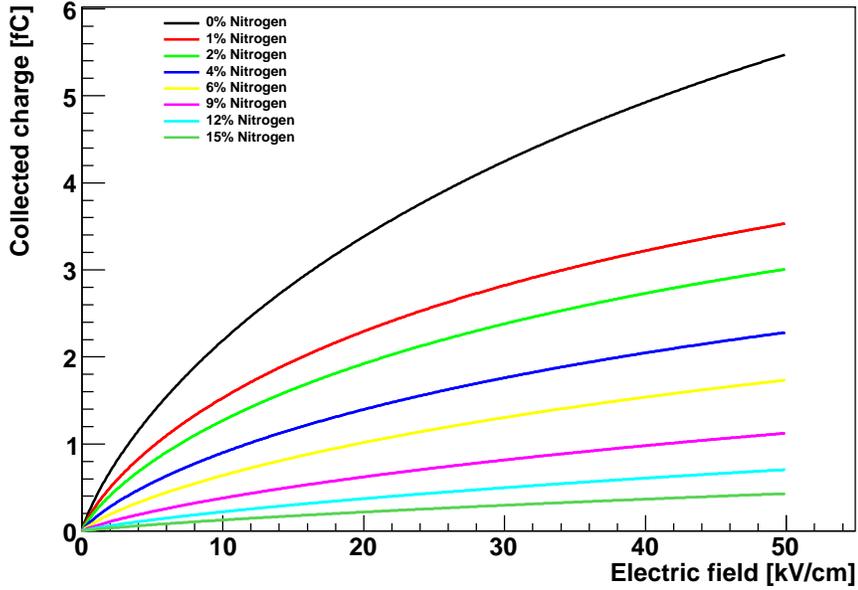}
\caption{Expected ionization yield from protons of 1.75 MeV in a liquid $Ar-N_2$ TPC.}\label{fig:proton}
\end {figure}

By using the obtained dependences of the $W$ and $\kappa$ parameters on the Nitrogen concentration, charge-field curves for 1.75~MeV GRNA protons similar to those of Figures \ref{fig:AlphaG8}, \ref{fig:AlphaG5} and \ref{fig:Gspec} can be predicted. These curves for different Nitrogen concentrations are shown in Figure \ref{fig:proton}.
The derivation of these dependencies is the main goal of the presented work and gives a hint about the operating conditions of the liquid $Ar-N_2$ TPC in order to be used as a selectively sensitive detector for GRNA radiography.
With the charge readout sensitive to ionization signals of 0.2~fC at a drift field of 50~kV/cm the content
of Nitrogen in Argon can be as high as 15\% by volume. These results indicate that liquid $Ar-N_2$ TPC detectors
can be fruitfully employed for GRNA applications.

\section{Conclusions}

A short drift gap Time Projection Chamber (TPC) filled with a mixture of liquid Argon and Nitrogen has been realized and successfully operated with Nitrogen concentrations of up to 15\% for the first time. The performance of the TPC has been studied under different experimental conditions, following a first set of measurements discussed in \cite{nitroTPC}. 

A precise measurement of the Nitrogen concentration is a key issue for the energy calibration of the TPC. For this reason, a specific resonance-based density meter was developed and an accuracy of the order of 0.3\% was obtained.

Detailed studies on electron-ion recombination for different $Ar-N_2$ mixtures with $^{60}Co$-Compton electrons and alpha-particles from $^{241}Am$ were performed. The so-called Box model describing the electron-ion recombination is used to fit the data, assuming that both the charge yield $W$ and the recombination parameter $\kappa$ depend on the Nitrogen concentration. These are the first measurements
of the ionization charge behaviour for both light and heavy ionizing particles in the mixtures of the liquid Argon and Nitrogen with Nitrogen concentrations up to 15\%.

The experimental results obtained for electrons and alpha-particles were combined to derive the expected ionization characteristics  for 1.75~MeV protons. These results are very important to define the optimal working conditions for Gamma Resonant Nuclear Absorbtion (GRNA) radiography. 

The results of these measurements are encouraging for future developments of liquid $Ar-N_2$ TPCs. However, more studies will be needed in view of the application to the Nitrogen GRNA radiography. In particular, the study of the scintillation signal provided by particles with different ionization densities can give additional information for proton-electron identification in liquid $Ar-N_2$ TPCs. For the future we plan to continue R\&D work along this direction, towards the development of a working GRNA detector based on this principle.

\section*{Acknowledgements}
We wish to thank all our technical collaborators for their very valuable help in building and operating the detector and the related infrastructure, namely S.~Lehman, F.~Nydegger, J.~Christen, R.~Mathieu. We would also express gratitude to our engineer staff: R.~H\"anni, P.~Lutz,  L.~Martinez.
Finally we are indebted to Dr.~S.~Braccini for reviewing the paper and providing valuable suggestions.


\begin{thebibliography}{99}


\bibitem{Rubbia77} C. Rubbia, \emph{The Liquid Argon Time Projection Chamber: a New Concept for Neutrino Detector}, CERN-EP (1977) 77-08.

\bibitem{Attie} D. Attie, \emph{TPC review}, Nucl. Inst. Meth. A598 (2009) 89-93.

\bibitem{Ereditato1} A. Ereditato and A. Rubbia, \emph{Ideas for future liquid Argon detectors}, 
Nucl. Phys. Proc. Suppl. 139 (2005) 301.

\bibitem{Ereditato2} A. Ereditato and A. Rubbia, \emph{The liquid Argon TPC: a powerful detector for future neutrino experiments
and proton decay searches}, Nucl. Phys. Proc. Suppl. 154 (2006) 163.

\bibitem{Ereditato3} A. Ereditato and A. Rubbia, \emph{Conceptual design of a scalable multi-kton superconducting magnetized
liquid Argon TPC}, Nucl. Phys. Proc. Suppl. 155 (2006) 233.

\bibitem{nitroTPC}
A. Ereditato et al., \emph{Study of ionization signals in a TPC filled with a mixture of liquid Argon and Nitrogen},
JINST 3 : P10002, 2008.

\bibitem{Biagio} B. Rossi et al., \emph{A Prototype liquid Argon Time Projection Chamber for the study of UV laser multi-photonic ionization}, JINST 4:P07011. 

\bibitem{Biagio2} I. Badhrees et al., \emph{Measurement of the two-photon absorption cross-section of liquid Argon with a Time
Projection Chamber}, accepted for publication in New Journal of Physics, 2010

\bibitem{GRA1} M. Goldberg et al., \emph{A method for detection of explosives based on nuclear resonance absorption of gamma rays in $^{14}N$}, Nucl. Inst. Meth. A348 (1994) 688-691. 

\bibitem{GRA2} G. Feldman et al., \emph{Analysis of gamma-ray nuclear resonant absorption (NRA) images for automatic explosives detection}, Seventh International Conference on Image Processing and Its Applications, Vol. 2 (1999) 789-793.

\bibitem{GRA3} L. Wielopolski et al., \emph{Gamma resonance absorption. New approach in human body composition studies}, Annals of New York Academy of Sciences, 904 (2000) 229-235.

\bibitem{GRA4} D. Vartsky et al., \emph{Gamma Ray Nuclear Resonance Absorption: An Alternative Method for in Vivo Body Composition Studies},  Annals of the New York Academy of Sciences 904 (2000) 236-246.

\bibitem{DavidsonLarsh} N. Davidson and A. E. Larsh, Jr.,\emph{Conductivity Pulses in Liquid Argon}, Phys. Rev. 74 (1948) 220-220.

\bibitem{Swan} D. W. Swan, \emph{Electron Attachment Processes in Liquid Argon containing Oxygen or Nitrogen Impurity},  Proc. Phys. Soc. 82 (1963) 74-84.

\bibitem{Swan2} D. W. Swan, \emph{Drift velocities of electrons in liquid argon, and the influence of molecular impurities},
Proc. Phys. Soc. 83 (1964) 659-666.

\bibitem{Yoshino} K. Yoshino et al., \emph{Effect of molecular solutes on the electron drift velocity in liquid Ar, Kr and Xe}, Phys. Rev. A14 (1976) 438-444.

\bibitem{Barabash} A.S.Barabash et al., \emph{Investigation of electronic conductivity of liquid argon-nitrogen mixtures}, Nucl. Inst. Meth. A234 (1985) 451-454. 

\bibitem{Shibamura} E. Shibamura et al., Nucl. Inst. Meth. A131 (1975) 249.

\bibitem{Sakai}  T. Kimura et al., \emph{Fast and slow electrons in liquid Ar and N2 mixtures}, IEEE Transactions on Dielectrics and Electrical Insulation, Vol. 1, Issue 4 (1994) 644-647.

\bibitem{Sakai2}  Y. Sakai et al., \emph{Excess electrons in $N_2/Ar$ liquid mixtures}, Nucl. Inst. Meth. A327 (1993) 92-94. 

\bibitem{Marcel}
M. Zeller, \emph{Development of a TPC filled with a mixture of liquid Argon and Nitrogen for $\gamma$-ray resonant absorption applications}, Master-Thesis der Philosophisch-naturwissenschaftlichen Fakult\"at der Inivesit\"at Bern,
2009. 

\bibitem{ThomasImel} J. Thomas and D. A. Imel, \emph{Recombination of electron-ion pairs in liquid argon and liquid xenon}, Phys. Rev. A36 (1987) 614-616.

\bibitem{CSDA} M.J. Berger et al., \emph{Stopping-Power and Range Tables for Electrons, Protons, and Helium Ions}, 
http://www.nist.gov/physlab/data/star/index.cfm

    
\bibitem{Vuillemin} V. Vuillemin et al., \emph{Electron drift velocities and characteristics of ionization of alpha and beta particles in liquid Argon doped with ethylene for LHC calorimetry}, Nucl. Inst. Meth. A316 (1992) 71-82. 

\end{thebibliography}
\end{document}